\documentclass[12pt,preprint]{aastex}
\begin{document}
\newcommand{\up}[1]{\ifmmode^{\rm #1}\else$^{\rm #1}$\fi}
\newcommand{\zdot}{\makebox[0pt][l]{.}}
\newcommand{\upd}{\up{d}}
\newcommand{\uph}{\up{h}}
\newcommand{\upm}{\up{m}}
\newcommand{\ups}{\up{s}}
\newcommand{\arcd}{\ifmmode^{\circ}\else$^{\circ}$\fi}
\newcommand{\arcm}{\ifmmode{'}\else$'$\fi}
\newcommand{\arcs}{\ifmmode{''}\else$''$\fi}

\title{The Araucaria Project. Accurate determination of the dynamical mass of the 
classical Cepheid in the eclipsing system OGLE-LMC-CEP-1812\footnote{Based on observations obtained 
with the ESO VLT and 3.6 m  telescopes
for Programmes 084.D-0640(A), 085.D-0398(A), and 086.D-0103(A), and with the Magellan Clay telescope
at Las Campanas Observatory} }

\author{Grzegorz Pietrzy{\'n}ski}
\affil{Universidad de Concepci{\'o}n, Departamento de Astronomia,
Casilla 160-C, Concepci{\'o}n, Chile}
\affil{Warsaw University Observatory, Al. Ujazdowskie 4, 00-478, Warsaw, Poland}
\authoremail{pietrzyn@astrouw.edu.pl}
\author{Ian B. Thompson}
\affil{Carnegie Observatories, 813 Santa Barbara Street, Pasadena, CA,911101-1292}
\authoremail{ian@obs.carnegiescience.edu}
\author{Dariusz Graczyk}
\affil{Universidad de Concepci{\'o}n, Departamento de Astronomia,Casilla 160-C, Concepci{\'o}n, Chile}
\authoremail{darek@astro-udec.cl}
\author{Wolfgang Gieren}
\affil{Universidad de Concepci{\'o}n, Departamento de Astronomia,
Casilla 160-C, Concepci{\'o}n, Chile}
\authoremail{wgieren@astro-udec.cl}
\author{Bogumi{\l} Pilecki}
\affil{Warsaw University Observatory, Al. Ujazdowskie 4, 00-478, Warsaw,
Poland}
\authoremail{pilecki@astrouw.edu.pl}
\author{Andrzej  Udalski}
\affil{Warsaw University Observatory, Al. Ujazdowskie 4, 00-478, Warsaw, Poland}
\authoremail{udalski@astrouw.edu.pl}
\author{Igor  Soszynski} 
\affil{Warsaw University Observatory, Al. Ujazdowskie 4, 00-478, Warsaw, Poland}
\authoremail{soszynsk@astrouw.edu.pl}
\author{Giuseppe Bono}
\affil{INAF-Osservatorio Astronomico di Roma, Via Frascati 33, 00040 Monte Porzio Catone}
\authoremail{Giuseppe.Bono@roma2.infn.it}
\author{Piotr Konorski}
\affil{Warsaw University Observatory, Al. Ujazdowskie 4, 00-478, Warsaw, Poland}
\authoremail{piokon@astrouw.edu.pl}
\affil{Universidad de Concepci{\'o}n, Departamento de Astronomia, Casilla 160-C, Concepci{\'o}n, Chile}
\author{Nicolas Nardetto}
\affil{Laboratoire Fizeau, UNS/OCA/CNRS UMR6525, Parc            
Valrose, 06108 Nice Cedex 2, France}
\authoremail{Nicolas.Nardetto@oca.eu}
\author{Jesper Storm}
\affil{Leibniz Institute for Astrophysics, An der     
Sternwarte 16, 14482, Postdam, Germany}
\authoremail{jstorm@aip.de}

\begin{abstract}
We have analyzed the double-lined eclipsing binary system OGLE-LMC-CEP-1812 in
the LMC and demonstrate that it contains a classical fundamental mode Cepheid
pulsating with a period of 1.31 days. The secondary star is a stable giant. We
derive the dynamical masses for both stars with an accuracy of 1.5\%, making the
Cepheid in this system the second classical Cepheid with a very accurate
dynamical mass determination, following the OGLE-LMC-CEP-0227 system studied by
Pietrzynski et al. (2010). The measured dynamical mass agrees very well with
that predicted by pulsation models. We also derive the radii of both components
and accurate orbital parameters for the binary system. This new, very accurate
dynamical mass for a classical Cepheid will greatly contribute to the solution
of the Cepheid mass discrepancy problem, and to our understanding of the
structure and evolution of classical Cepheids.
\end{abstract}


\section{Introduction}
Classical Cepheids  have been key objects in the long-lasting efforts to measure
the extragalactic distance scale and to probe the predictions of stellar
evolution and stellar pulsation theory (e.g. Freedman and Madore 2010, Caputo et
al. 2005). Given the enormous importance of Cepheids for the determination of
the cosmic distance scale and cosmological parameters, it is of great importance
to fully understand these stars astrophysically.  One of the most nagging
problems in Cepheid research has been the difficulty to reliably determine their
masses. Christy (1968) and Stobie (1969) were the first to notice that Cepheid
masses calculated from stellar pulsational theory were about 20 \% smaller than
the corresponding masses estimated from their evolutionary tracks on the
Hertzsprung-Russell diagram. In spite of the considerable progress in
understanding the physics of Cepheid variable stars over the years, the "Cepheid
mass discrepancy problem" has remained unsolved for more than 40 years (Keller
and Wood 2002, Keller 2008, Evans et al. 2008, Caputo et al. 2005, Neilson,
Cantiello and Langer 2011 and references therein). The obvious solution to the
problem comes from an independent and accurate measurement of the {\it dynamical
masses} of a number of Cepheids with a range of pulsation periods, and there
have been a number of efforts to find binary Cepheids allowing such a precise
mass determination (Evans et al. 1997, 2006 and 2008). However, the few Cepheids
found in binary systems for which masses have been estimated all occur in single-lined,
non-eclipsing systems. This has limited the accuracy of the best Cepheid dynamical
mass measurements to less than 15~\%, not sufficiently accurate to resolve the
mass discrepancy problem.

Recently, several eclipsing binary systems in the LMC have been discovered from OGLE
Project data which contain candidates for classical Cepheid components (Udalski
et al. 1999, Soszynski et al. 2008). Our group, as part of the  {\it Araucaria
Project} dedicated to the improvement of stellar distance indicators (Gieren et al. 2005),
has been obtaining spectroscopic observations of these candidates to confirm the true
binary nature of these systems and the Cepheid nature of the intrinsically
variable components. Accurate high resolution spectra
collected by our group for  OGLE-LMC-CEP-0227 (hereinafter CEP-0227)
resulted in the first confirmed
discovery of a classical Cepheid variable in an eclipsing system, perfectly
suited for an accurate determination of  its physical parameters (Pietrzynski et
al. 2010).  The analysis of our data led to the determination of the dynamical
mass of this 3.8-day  Cepheid with an unprecedented accuracy of 1~\%.
This study provided strong constraints on theoretical
pulsation and evolutionary models and favored the mass predictions for Cepheids
from pulsation theory. Cassisi and Salaris (2011) have analyzed
evolutionary models of this Cepheid finding good agreement between its
evolutionary and dynamical mass when extra mixing is included.  Neilson et al. (2011)
find that the mass discrepancy can be accounted for if the evolutionary
models include convective mass overshooting and pulsation-driven mass loss.
The models also predict that the size of the Cepheid mass discrepancy depends on the Cepheid
mass, with the size of the mass discrepancy expected to rise to lower periods
( Caputo et al. 2005). On the other hand Keller (2008) showed that the predicted behavior of the mass 
discrepancy depends on the assumed mass-luminosity relation. Therefore it is clearly desirable 
to measure accurate dynamical masses
for several additional Cepheids spanning a range in period in order to fully
resolve this problem and to understand the structure, physics and evolution of
Cepheid variables.

In this paper we present the analysis of the second confirmed classical Cepheid
in a double-lined eclipsing binary system, OGLE-LMC-CEP-1812
(hereinafter CEP-1812), and derive its
dynamical mass with an accuracy of 1.5 \%. Our new binary Cepheid has a shorter
pulsation period of 1.31 days compared to 3.80 days for CEP-0227, and its measured 
physical properties therefore present a valuable complement to those derived 
for CEP-0227.

\section{Observations}

\subsection{Optical Photometry}
The optical phtometry of CEP-1812 was obtained with the Warsaw
1.3m telescope at Las Campanas Observatory in the course of the
third phase of the OGLE project (e.g. Soszynski et al. 2008).
A total of 883 $I$-band  epochs spanning a period of 1076 days were
secured.  The data were reduced with the
image-subtraction technique (Udalski 2003, Wozniak 2000).
The instrumental data were 
calibrated onto the standard system using observations of several
Landolt fields over several photometric nights. The estimated 
zero point errors are about 0.01 mag  (Udalski et al. 2008).
For more details about the instrumental system, observing, reduction and
calibration procedures adopted in the course of the OGLE project
the reader is referred to the references cited above.  

\subsection{High Resolution Spectroscopy}
High resolution spectra of the CEP-1812 system were collected with the
Las Campanas Observatory Magellan Clay 6.5~m telescope and the MIKE
echelle spectrograph (Bernstein et al. 2003), with the ESO VLT Kueyen 
8.2~m telescope and the UVES echelle spectrograph (Dekker et al. 2000),
and with ESO 3.6~m telescope and the HARPS fiber-fed echelle spectrograph
(Mayor et al. 2003).
In the case of the MIKE observations a 0.7 arcsec slit was used
giving  a  resolution of  about 40,000. The spectra were reduced with 
pipeline software developed by Kelson (2003). Exposure times ranged from 2400~sec to
3200~sec depending on observing conditions, and a typical resulting S/N ratio 
was 10 at a wavelength of 4500~\AA. The UVES observations were obtained with an
exposure time of 2700~sec  and a 0.7 arcsec slit resulting in a spectral
resolution of about 50,000 and a S/N ratio at 4500~\AA~of better than 15. The
UVES data were reduced with the ESO pipeline. The HARPS observations were obtained at a
resolution of 60,000 and a S/N at 5000~\AA~of 4 for one hour
integrations, and were reduced with the data reduction software
deweloped by the Geneva observatory. In total 88 high quality spectra
were collected with the three spectrographs.  

CEP-1812 is located in a dense
region in the LMC and even based on inspection of the OGLE template images 
with a spatial resolution of 0.8 arcsec (Soszynski et al. 2008)  
one can see several faint stars located very close to the star. Simple cross correlation
measurements showed the presence of a third velocity component.
We therefore decided to measure velocities for CEP-1812 with the broadening function (BF)
formalism (Rucinski 1992), which is widely recognized as an effective
technique for radial velocity determination from complex spectra
(Rucinski 2003). The BFs were calculated on the wavelength
interval 4350~\AA~to 6100~\AA~using a template interpolated from the
Coelho library of synthesized echelle resolution spectra (Coelho 
et al. 2005). The BFs revealed  clear peaks for both
components of CEP-1812 in all spectra. In addition, the third  signal was detected in
some 40~\% of our spectra at a constant
radial velocity of 273 $\pm$ 5 km/s. The BF signal of 
this star was used to  calculate its light contribution to the total
I-band brightess of the system to be  about 10~\%. We found no significant 
offsets between the velocity systems of the three instruments.

\section{Spectroscopic and Photometric Solutions}

Based on  the photometric data the following ephemerides were derived:

$ {\rm P_{orb}}  =  551.798 \pm  0.010 $ days $ \hspace*{2cm}  {\rm T}_{\rm 0, orb} = 2450479.11 \pm 0.07 $ \\
$ {\rm P_{pul}}  =  1.312904 \pm  0.000003 $ days $ \hspace*{2cm}  {\rm T}_{\rm 0, pul} = 2450455.685 \pm 0.006 $ \\

Adopting these ephemerides the spectroscopic orbit (systemic velocity, velocity
amplitudes, eccentricity, periastron passage and mass ratio) plus a Fourier
series of order six (which approximates the pulsations of the Cepheid primary
component) were fitted to the radial velocity data. The orbit solution
and the pulsational radial velocity curves of the Cepheid component are shown in
Figure 1. Figure 2 shows the pulsational $I$-band light curve of the Cepheid.
Adopting the obtained spectroscopic mass ratio of 0.705 $\pm$ 0.015,
and an I-band third light contribution of 10 \%, we modeled the  
OGLE-LMC-CEP-1812 system with the
Wilson Devinney code (Wilson and Devinney 1971, Van Hamme and Wilson, 2007) in
an iterative way removing the intrinsic brightness variations of the Cepheid
component.  The parameters corresponding to our best model together with their
uncertainties estimated from extensive Monte Carlo simulations are presented in
Table 1. Unfortunately, at the present moment without near infrared photometry,
we are not in a position to derive ${\rm T}_{\rm eff}$ of the components. 
Figure 3 presents the light curve of the eclipsing system with the
intrinsic brightness variations of the Cepheid removed, together with our best fit
WD model for the system.  Although the relative distance between components is
very large the eclipses are total with the secondary transiting over the Cepheid
disk during the primary eclipse. It is worth mentioning that the presence of
the third component only marginally affects the mass determination. Neglecting
the third light leads to an inclination of 89.5 degrees.

\section{Discussion and Conclusions}
The Cepheid in OGLE-LMC-CEP-1812 is a classical Cepheid, and not a Type II low-mass
Cepheid. The evidence comes from its mass of 3.74 $\pm$ 0.06
$M_{\bigodot}$ which agrees well with the predicted pulsational mass for a
classical Cepheid of this short period (Bono et al. 2001), and from its radius
(17.4 $\pm$ 0.9 $R_{\bigodot}$) which again is in good agreement with the radius
of a classical Cepheid predicted from period-radius relations (e.g. Gieren et
al. 1998). The position of the Cepheid in the period-mean magnitude plane for
LMC Cepheids shown in Figure 4 also proves beyond any doubt that CEP-1812 is a
classical Cepheid. Based on the analysis of the Fourier parameters Soszynski et al. (2008) 
firmly classified this star as a fundamental mode pulsator.
The system OGLE-LMC-CEP-1812 is thus the second known double-lined
eclipsing binary system with a classical fundamental mode Cepheid component. The
secondary component is a less massive, smaller and cooler stable giant star, a
configuration which is significantly different from the OGLE-LMC-CEP-0227 system
whose stable secondary giant star has the same mass, and a slightly larger
diameter than the Cepheid in that system. The configuration is also quite unusual
from an evolutionary point of view because the system consists of two well separated stars in
a relatively short stage of common giant phase evolution in spite of the large mass
difference. According to the BaSTI evolutionary tracks (Pietrinferni et al. 2004)
a star with mass 3.74 $M/M_{\bigodot}$ will enter the instability strip from the
giant branch after approximately 190 Myr of evolution, whereas a star of mass 2.64 $M/M_{\bigodot}$
will reach a radius of 12.1 $R/R_{\bigodot}$ on the subgiant branch after
approximately 369 Myr of evolution. While detailed evolutionary models of the 
two stars in the binary will be needed, these estimates call into question the
assumption that the stars are members of a coeval binary system.

In order to calculate the pulsational mass of CEP-1812 we adopted a
period-mass-relation  based on nonlinear, convective Cepheid models constructed
for the typical chemical composition of LMC Cepheids (Z(metals)=0.008,
Y(helium)=0.256) (Bono et al.~2000, Luck et al. 1998). We note that the
calculation of the  pulsation mass depends neither on the assumed
distance nor on the reddening. The resulting pulsation mass (3.27 $\pm$ 0.64
$M_{\bigodot}$) agrees very well with the dynamical mass of the star.
Pietrzynski et al. (2010)  came to the same conclusion for CEP-227. Therefore,
we have now strong observational evidence that the pulsation mass of a Cepheid
variable is indeed correctly measuring its true, current mass.

The dynamical mass determinations for both stars in OGLE-LMC-CEP-1812 are
accurate to about 1.5\%, adding them to the very short list of evolved massive
stars with very accurate mass determinations. In a forthcoming study, we will
discuss  evolutionary models for the Cepheid in the system which
will further add to the solution of the Cepheid mass discrepancy problem, and to
a deeper understanding of the physics and evolution of classical Cepheid
variable stars.

\acknowledgments
We gratefully acknowledge financial support for this
work from the Chilean Center for Astrophysics FONDAP 15010003, and from
the BASAL Centro de Astrofisica y Tecnologias Afines (CATA) PFB-06/2007. 
Support from the Polish grant N203 387337, and the FOCUS 
and TEAM subsidies of the Fundation for Polish Science (FNP)
is also acknowledged. IBT acknowledges the support of NSF grant AST-0507325.
The OGLE project has received funding from the European Research Council        
under the European Community's Seventh Framework Programme                      
(FP7/2007-2013) / ERC grant agreement no. 246678.
It is a pleasure to thank the support staff at ESO-Paranal and at Las Campanas Observatory
for their expert help in obtaining the observations. We also thank the ESO OPC and CNTAC
for allotting generous amounts of observing time to this project.

\begin{figure}[p] 
\vspace*{21 cm}   
\includegraphics{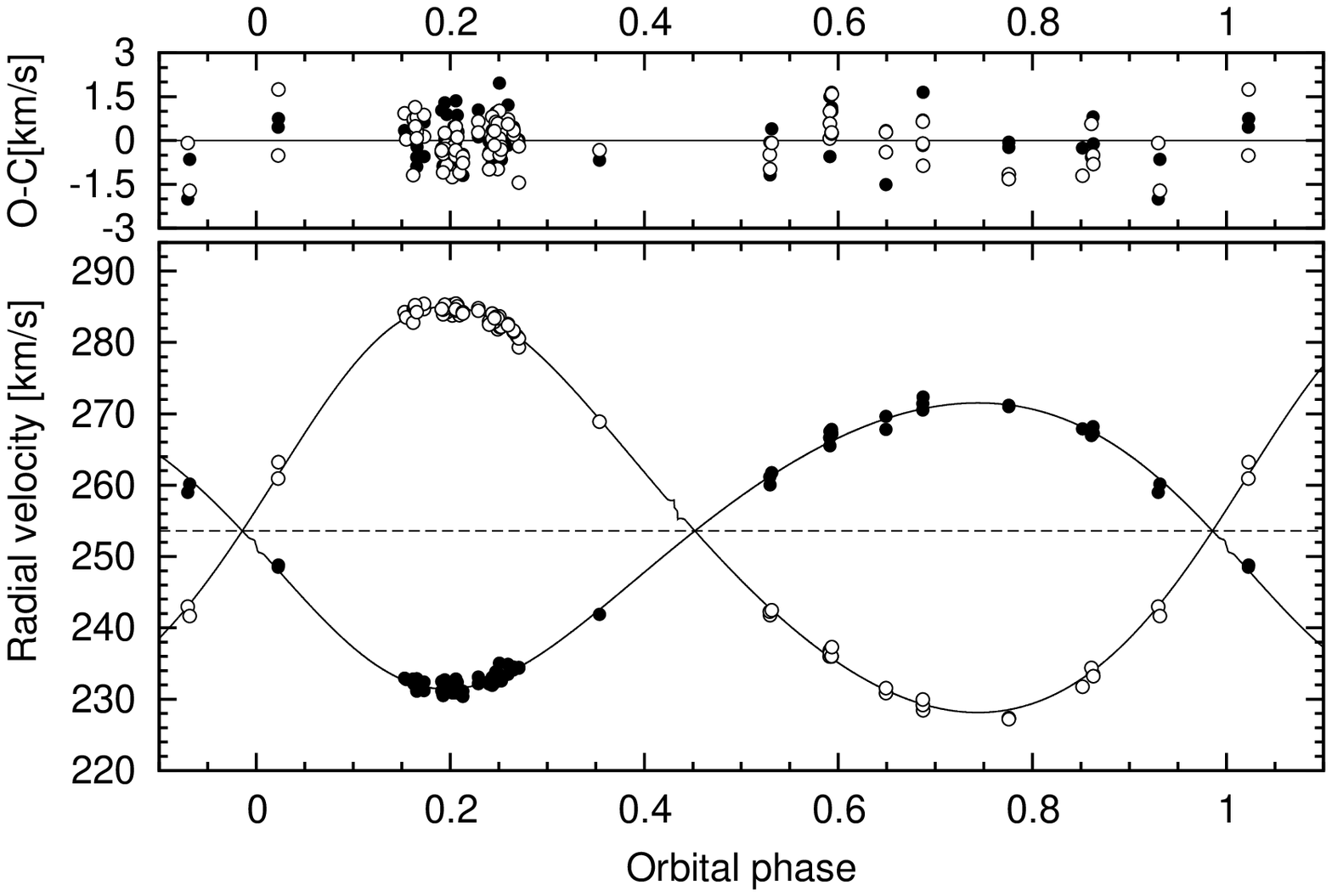}
\includegraphics{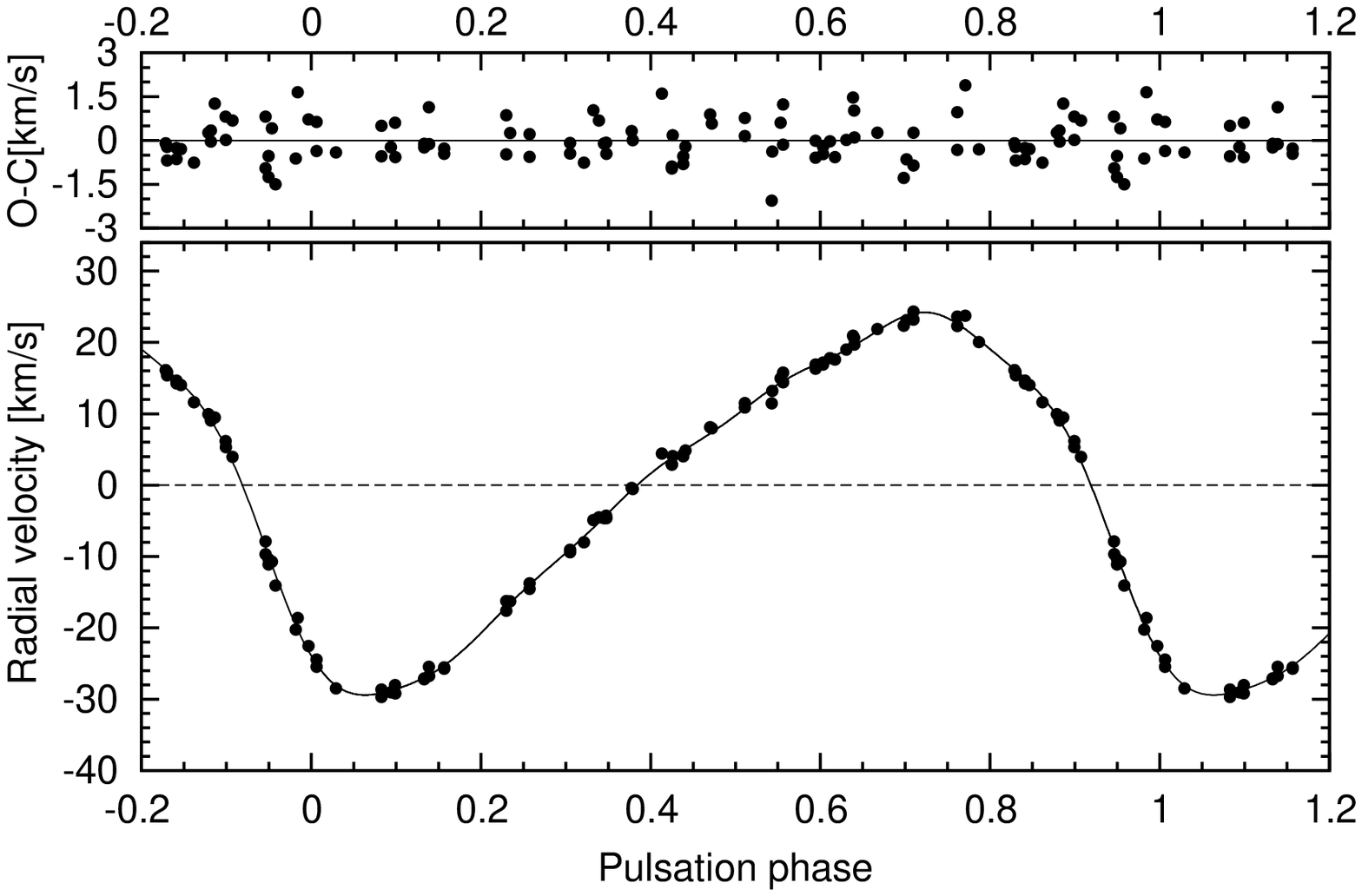}
\caption{Disentangled spectroscopic orbit of the CEP-1812 system, folded on the
orbital period of 551.8 days (upper panel), and the pulsational radial velocity curve
of the Cepheid, folded on its pulsation period of 1.31 days  (lower panel). The Cepheid 
is the higher-mass star in the system (filled circles in the upper panel). The measured  
(constant) velocities of the faint third star in the light are not shown for clarity. }
\end{figure}  

\begin{figure}[p]
\vspace*{12 cm}  
\includegraphics{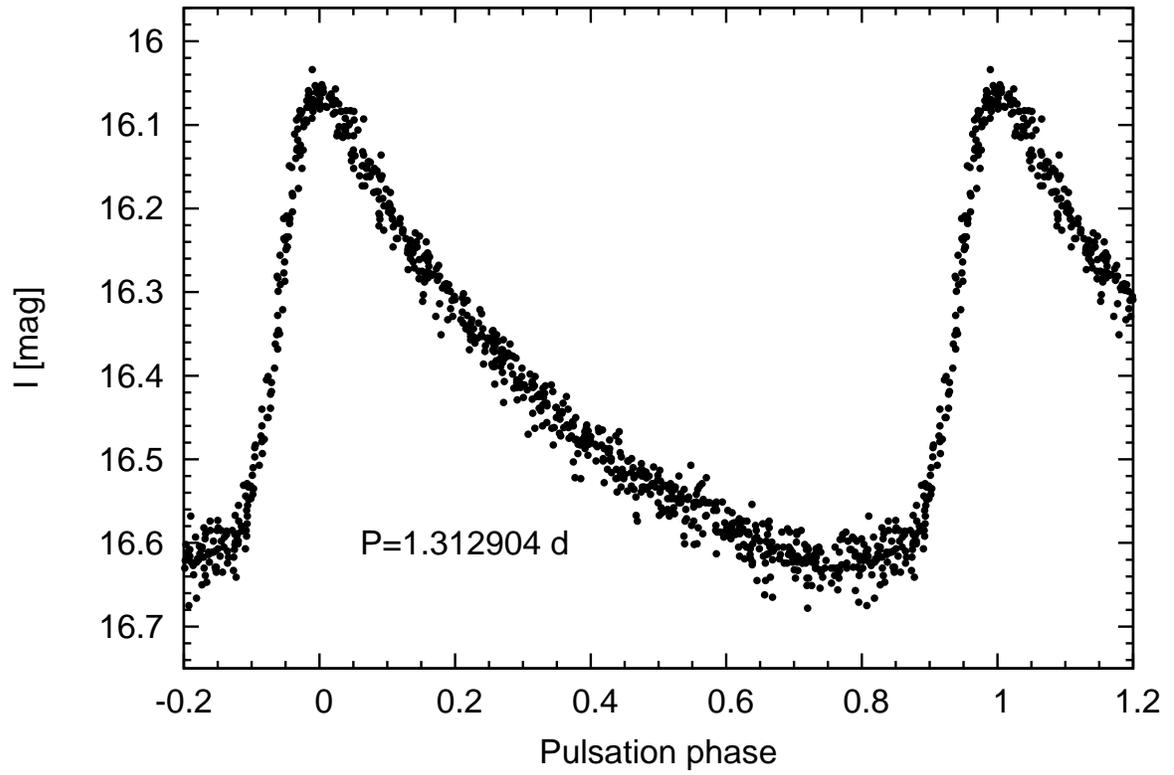}
\caption{The disentangled photometric $I$-band light curve of the Cepheid component in CEP-1812
from 883 individual measurements.}
\end{figure}

\begin{figure}[p]
\vspace*{22 cm}
\includegraphics{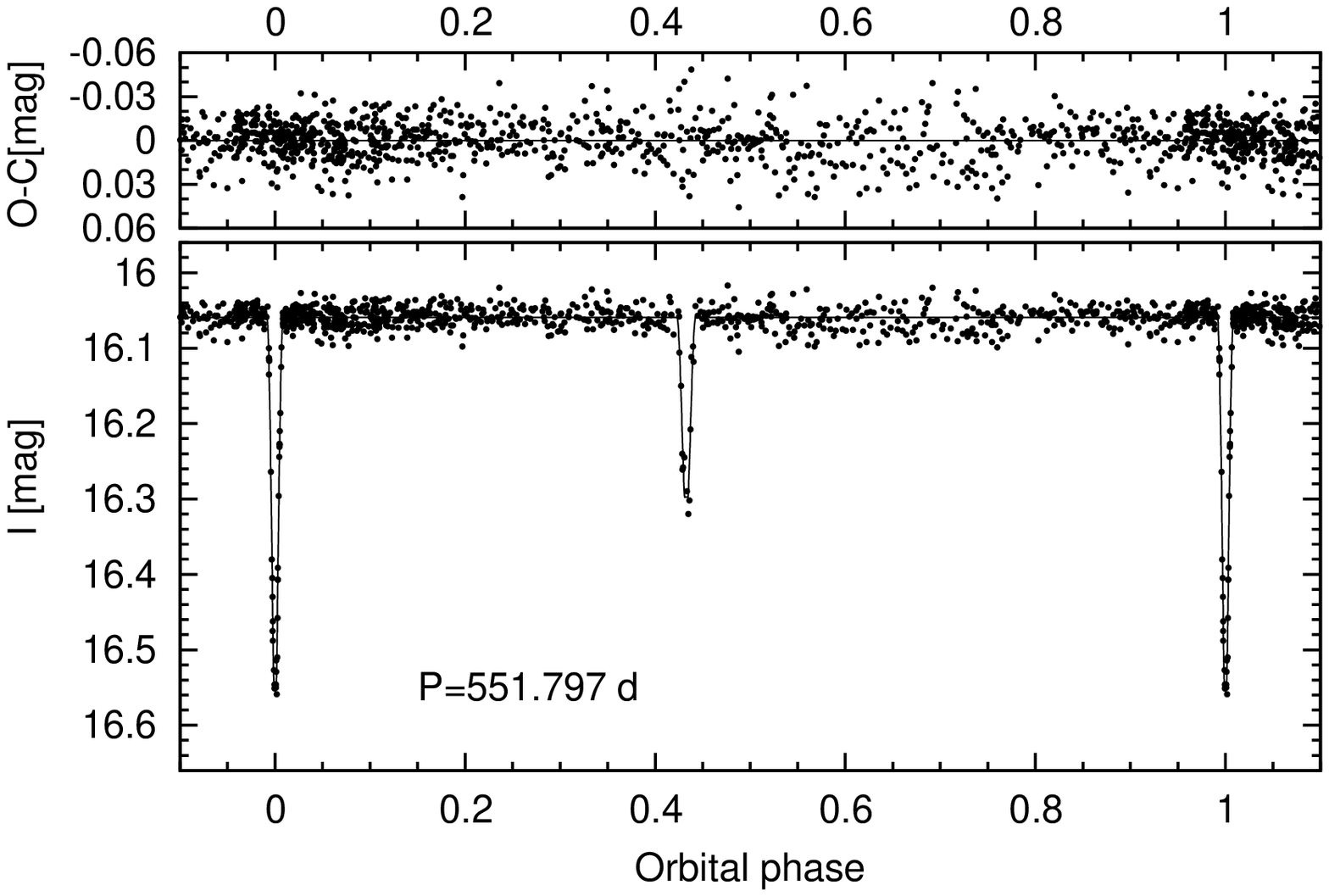}
\includegraphics{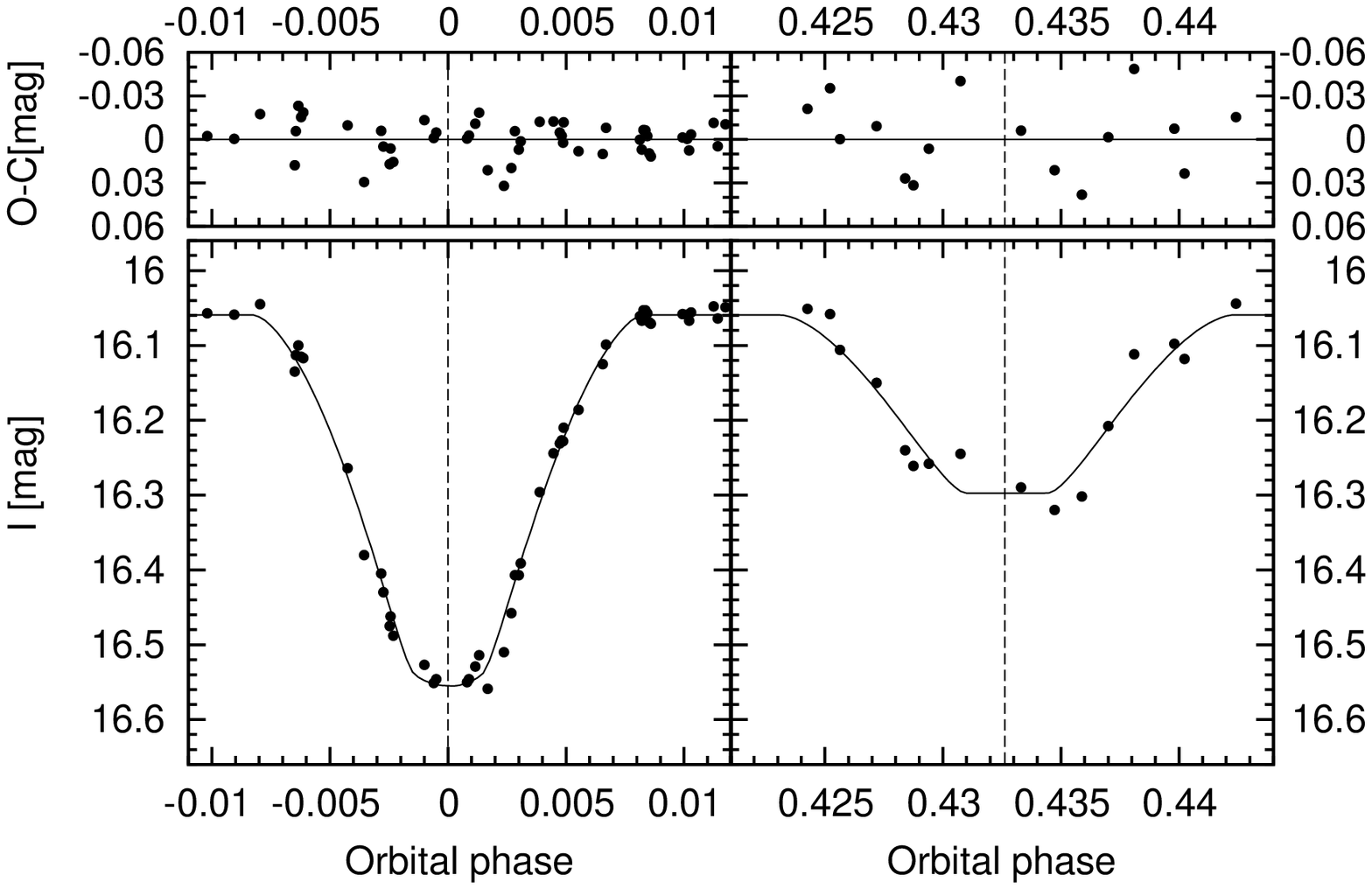}
\caption{Observed orbital $I$-band light curve together with the photometric solution
as obtained from the analysis with the Wilson-Devinney code. 
}
\end{figure}

\begin{figure}[p]
\vspace*{12 cm}  
\includegraphics{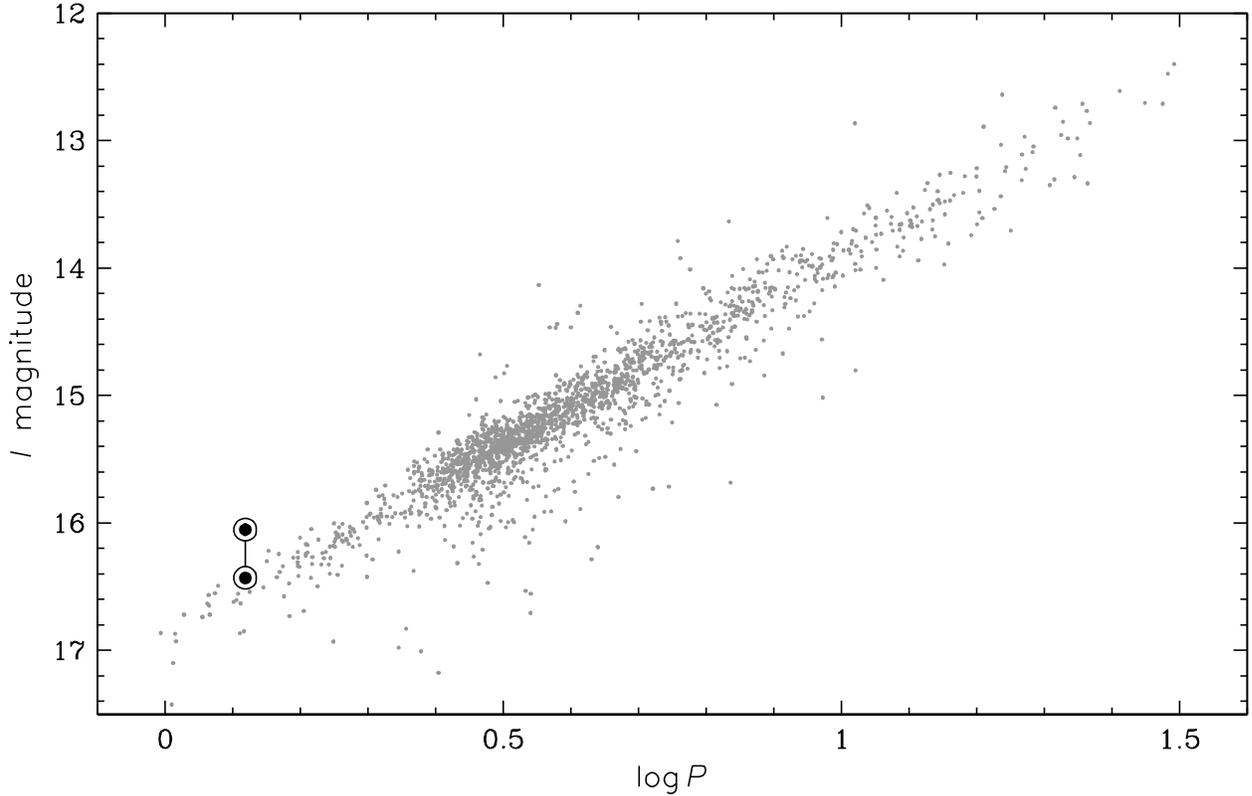}
\caption{The $I$-band period-luminosity relation (P in days) for classical
fundamental mode Cepheids as defined by the catalog of OGLE III Cepheids in the
LMC (Soszynski et al. 2008). The position of our binary system is marked by
the upper filled circle (out-of-eclipse total brightness of the system, including the faint
third star in the light), and the position of CEP-1812 (its mean brightness,
freed from the contributions coming from the binary companion and the faint third star) is indicated 
by the lower circle. The Cepheid falls almost exactly on the ridge line of the
period-magnitude relation indicating that CEP-1812 is a classical Cepheid.
}
\end{figure}

\begin{deluxetable}{cc}
\tablecaption{Astrophysical parameters of the OGLE-LMC-CEP1812 system. 
The quoted uncertainties were estimated from extensive Monte Carlo simulations.}
\startdata
 P (orbital) [days]         & 551.797 $\pm$ 0.010  \\
P (pulsational) [days]         & 1.312904 $\pm$ 0.000003 \\
 i [deg]          &     90.0 $\pm$ 0.4   \\ 
 a [$R_{\bigodot}$] &     524.5 $\pm$ 1.1  \\
 e                & 0.129 $\pm$ 0.012  \\
 $\omega$  [deg]  & 144.9 $\pm$ 4.7  \\
 $ q = m_{2} / m_{1}$ & 0.705 $ \pm $ 0.015  \\ 
 $ \gamma$ [km/s]        &  253.6 $\pm 0.3 $  \\
$M/M_{\bigodot}$ (Cepheid)  &  3.74 $\pm$ 0.06   \\
$M/M_{\bigodot}$ (secondary component)   &  2.64 $\pm$ 0.04 \\
$R/R_{\bigodot}$ (Cepheid)  & 17.4 $\pm$ 0.9   \\
$R/R_{\bigodot}$ (secondary component) & 12.1 $\pm$ 2.3 \\
L2/L1 (I) & 0.28 $\pm$  0.10 \\
L3 (I)    & 0.1 * (L1+L2) \\
\enddata
\end{deluxetable}


\begin{references}
\reference{} Bernstein, R., Shectman, S. A., Gunnels, S. M., Mochnacki, S. \& Athey, A. 2003, SPIE, 4841, 1694
\reference{} Bono, G., Castellani, V., \& Marconi, M. 2000, \apj, 529, 293
\reference{} Bono, G., Gieren, W., Marconi, M., Fouqu{\'e}, P., \& Caputo, F. 2001, \apj, 563, 319
\reference{} Caputo, F., Bono, G., Fiorentino, G., Marconi, M., \& Musella, I. 2005, \apj, 629, 1021
\reference{} Cassisi, S., \& Salaris, M. 2011, \apj, 728, L43
\reference{} Christy, R.F. 1968, QJRAS, 9, 13
\reference{} Coelho, P., Barbuy, Melendez, J., Sciavon, R.P., \& Castilho, B.V. 2005, \aap, 443, 735
\reference{} Decker, H. et al. 2000, SPIE, 4008, 534
\reference{} Evans N.R., Bohm-Vitense, E., Carpenter, K., Beck-Winchaz, B., \& Robinson, R. 1997, \pasp, 109, 789
\reference{} Evans, N.R., Massa, D., Fullerton, A., Sonneborn, G., \& Iping, R. 2006, \apj, 647, 1387
\reference{} Evans, N.R., Schaefer, G.H., Bond, H.E., Bono, G., Karovska, M., Nelan, E., Sasselov, D., \& Mason, B.D. 2008, \aj, 136, 1137
\reference{} Freedman, W.L., \& Madore, B.F. 2010, ARAA, 48, 673
\reference{} Gieren, W., Fouqu{\'e}, P., \& Gomez, M. 1998, \apj, 496, 17
\reference{} Gieren, W., Pietrzy{\'n}ski, G., Bresolin, F., et al. 2005, ESO Messenger, 121, 23
\reference{} Keller, S.C. 2008, \apj, 677, 483
\reference{} Keller, S.C., \& Wood, P.R. 2002, \apj, 578, 144
\reference{} Kelson, D. D. 2003, \pasp, 115, 688
\reference{} Luck, R.E., Moffett, T.J., Barnes, T.G., \& Gieren, W. 1998, \aj, 115, 605
\reference{} Mayor, M. et al. 2003, The Messenger, 114, 20
\reference{} Neilson, H.R., Cantiello, M., \& Langer, N. 2011, \aap, 529,L9
\reference{} Pietrinferni,, A., Cassisi, S., Salaris, M., \& Castelli, F. 2004, \apj, 612, 168 
\reference{} Pietrzy{\'n}ski, G., Thompson, I.B.,Graczyk, D., Gieren,
W., Udalski, A., Szewczyk, O., Minniti, D., Kolaczkowski, Z., Bresolin,
F., \& Kudritzki, R.P. 2009, \apj, 697, 862
\reference{} Pietrzy{\'n}ski, G., Thompson, I.B., Gieren, W., Graczyk, D., Bono, G., Udalski, A., Soszy{\'n}ski, I., Minniti, D., \& 
Pilecki, B. 2010, Nature, 468, 542
\reference{} Rucinski, S.M. 1992, \aj, 104, 1968
\reference{} Rucinski, S.M. 2003, Proceedings of IAU Symposium No. 215,
held 11-15 November, 2002 in Cancun, Yucatan, Mexico, 2004, p. 17 
\reference{} Soszy{\'n}ski, I., Poleski, R., Udalski, A., et al. 2008, Acta Astron., 58, 163
\reference{} Stobie, R.S. 1969, \mnras, 144, 511
\reference{} Udalski, A., Szymanski, M., Kubiak, M., Pietrzynski, G., Soszynski, I., Wozniak, P., 
\& Zebrun, K. 1999, Acta Astron., 49, 201
\reference{} Udalski, A., Soszynski, I., Szymanski, M., Kubiak, M., Pietrzynski, G., Wyrzykowski, 
{\L.}, Szewczyk, O., Ulaczyk, K., \& Poleski, R. 2008, Acta Astron., 58, 4632
\reference{} Udalski, A. 2003, Acta Astron., 53, 291
\reference{} Van Hamme, W., \&  Wilson, R.E. 2007, \apj, 661, 1129
\reference{} Wilson, R.E., \& Devinney, E.J. 1971, \apj, 166, 606
\reference{} Wo{\'z}niak, P. 2000, Acta Astron., 50, 421
\end{references}
\end{document}